# Transition state theory: a generalization to nonequilibrium systems with power-law distributions

Du Jiulin

*Department of Physics, School of Science, Tianjin University, Tianjin 300072,China*

Transition state theory (TST) is generalized for the nonequilibrium system with power-law distributions. The stochastic dynamics that gives rise to the power-law distributions for the reaction coordinate and momentum is modeled by the Langevin equations and corresponding Fokker-Planck equations. It is assumed that the system far away from equilibrium has not to relax to a thermal equilibrium state with Boltzmann-Gibbs distribution, but asymptotically approaches to a nonequilibrium stationary-state with power-law distributions. Thus, we obtain a generalization of TST rates to nonequilibrium systems with power-law distributions. Furthermore, we derive the generalized TST rate constants for one-dimension and $n$-dimension Hamiltonian systems away from equilibrium, and receive a generalized Arrhenius rate for the system with power-law distributions.

## I. INTRODUCTION

Transition state theory (TST) made it possible to obtain quick estimates for reaction rates of a broad variety of processes in physics, chemistry, biology, and engineering. It has been a cornerstone or a core of reaction rate theory and profoundly influenced the development of theory of chemical dynamics. However, one key assumption to TST is that thermodynamic equilibrium must prevail throughout the entire system studied for all degrees of freedom; all effects that result from a deviation from thermal equilibrium distribution, such as Boltzmann-Gibbs (B-G) distribution with the exponential law, are neglected [1]. In the reaction rate theory, we are interested in the processes of evolution from one metastable state to another neighboring state of metastable equilibrium; the assumption thus would be quite farfetched. An open problem, which is being investigated intensively, is reaction rate theory away from equilibrium. TST is no longer valid and cannot even serve as a conceptual guide for understanding the critical factors that determine rates away from equilibrium [2]. Thereby TST must be generalized if it is employed to describe rates of reactions in the nonequilibrium systems with non-exponential or power-law distributions.

Theoretically, if one supposes to generalize TST in calculating the rates of reactions in complex systems far away from thermal equilibrium, an approach may be to find stationary nonequilibrium distribution. But, because in most systems away from equilibrium the asymptotic probability distribution is not known due to the lack of detailed balance symmetry, it seems to be a quite formidable task to determine the probability density of multidimensional stationary nonequilibrium systems [2]. Often, numerical simulations [3] of the stochastic dynamics are the only possible approach. It is worth to notice that all the attempts so far ended up in having a form with the exponential law in B-G distribution. The need of generalizing TST rates in such a way as to have a non-exponential or power-law expression is due to the condition away from equilibrium since this condition creates genuine power-law distributions that have been noted prevalently, for example, in glasses [4,5], disordered media [6-8], folding of proteins [9], single-molecule conformational dynamics [10,11], trapped ion reactions [12], chemical kinetics, and biological and ecological population dynamics [13,14], reaction-diffusion processes [15], chemical reactions [16], combustion processes [17], gene expressions [18], cell reproductions [19], complex cellular networks [20], and small organic molecules [21] etc. In these situations, TST is invalid and the reaction rate equations have not existed yet.

On the other hand, a type of statistical mechanical theory of power-law distributions has been constructed via generalizing Gibbsian theory for systems away from thermal equilibrium [22]. Especially worth to mention that in recent years, nonextensive statistical mechanics based on the $q$-entropy proposed by Tsallis has received great attention and it has made very wide applications for a variety of interesting problems on those systems with power law distributions [23]. It is studied as a reasonable generalization of B-G statistical mechanics able to perform statistical description of a nonequilibrium stationary-state in the interacting systems [24-26], so becoming a very useful tool to approach complex systems whose properties go beyond the realm governed by B-G statistical mechanics. These developments now naturally give rise to a possibility for us to generalize TST to the nonequilibrium systems with power-law distributions, which is just our purpose in this work.

The paper is organized as follows. In section II, we deal with a type of Langevin equations for the reaction coordinate and momentum and the corresponding Fokker-Planck equations, and we study the conditions under which those equations will give rise to the power-law distributions. In section III, we study the TST for the nonequilibrium Hamiltonian systems with power-law distributions, including a generalization of TST rate constants to a one-dimension nonequilibrium system, to an *n*-dimension nonequilibrium system, and finally we derive the generalized Arrhenius rate for the power-law distributions. In section IV, we give the conclusion.

## II. STOCHASTIC DYNAMICS UNDERLYING POWER-LAW DISTRIBUTIONS

Let us consider the following reaction processes in an open system far away from equilibrium,

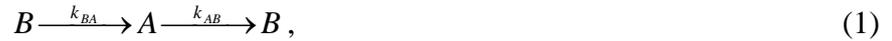

$$B \xrightarrow{k_{BA}} A \xrightarrow{k_{AB}} B, \tag{1}$$

where $k_{BA}$ is a rate constant of the reaction process changing the reactant $B$ into the product $A$, and $k_{AB}$ is that of the inverse reaction. If $N_A(t)$ and $N_B(t)$ denote the concentration of $A$ and $B$ at time $t$, respectively, the dynamical rate equation governing evolution of the instantaneous concentration is written as

$$\frac{dN_A(t)}{dt} = k_{BA} N_B(t) - k_{AB} N_A(t), \tag{2}$$

$$\frac{dN_B(t)}{dt} = k_{AB} N_A(t) - k_{BA} N_B(t). \tag{3}$$

These reaction processes with their interactions with the environment constitute a nonequilibrium dynamical reaction system. As we know that the stochastic motion of the reaction coordinate $x(t)$ is a combined effect induced by coupling among a multitude of environmental degrees of freedom. Conventionally in reaction rate theory, one assumes the reaction coordinate to describe dynamics of the escape process or the transition state, and its equation of motion to be modeled by the Langevin equation, where the system studied is regarded as ergodic [27] and the stationary macroscopic dynamics must have recourse to B-G statistical distribution [1]. However, for a complex system far away from equilibrium, the dynamics is not ergodic in general, and the stationary probability has not to be B-G distribution, but in many situations shows power-law distributions.

To investigate the stochastic dynamics underlying the power-law distributions in a system away from equilibrium, we start with the following Langevin equation for the dynamics of stochastic variables $x$,

$$\frac{dx}{dt} = K(x) + \eta_x(t), \tag{4}$$

with

$$\langle \eta_x(t) \rangle = 0, \quad \langle \eta_x(t)\eta_x(t') \rangle = 2D(x)\delta(t-t'), \tag{5}$$

where $K(x)$ is a given function of the coordinate $x$, It is required that the noise $\eta_x(t)$ is Gaussian distribution, with zero mean and the delta-correlated. The quantities $D(x)$ in general can be a function of $x$. Eq.(4) is quite a general Langevin equation with arbitrary function $K(x)$, very similar to a strong friction or an overdamped motion of interacting particles if $K(x)$ is associated with a potential $V(x)$ by $K(x) = -dV(x)/dx$ [1].

Corresponding to the dynamics governed by above Langevin equation (4), the Fokker-Planck (F-P) equation for the noise-averaged probability distribution function $\rho(x,t)$ [28] can be written by

$$\frac{\partial \rho(x,t)}{\partial t} = -\frac{\partial}{\partial x}[K(x)\rho(x,t)] + \frac{\partial}{\partial x}\left[D(x)\frac{\partial}{\partial x}\rho(x,p,t)\right]. \tag{8}$$

In this sense, $K(x)$ is usually the drift term and $D(x)$ the diffusion coefficient. It is easy to find that there exists a general stationary-state solution for the F-P equation (8), which can be expressed in an exponential form,

$$\rho_s(x) \sim \exp\left(\int dx [K(x)/D(x)]\right), \tag{9}$$

and $\rho_s(x) \to 0$ when $x \to \pm\infty$. Usually, in the simplest case, if we take $D(x)$ as a constant, e.g. $D=1/\beta$, and $K(x)$ as $K(x)= - dV(x)/dx$ in terms of an potential function $V(x)$, the stationary solution is Boltzmann-Gibbs distribution, $\rho_s(x) \sim \exp[-\beta V(x)]$, where $\beta = 1/kT$. However, for a given potential function $V(x)$, it is readily found that if the following condition is satisfied for the two any functions $K(x)$ and $D(x)$, and a given parameter $\kappa \neq 0$,

$$\frac{K(x)}{D(x)} = -\frac{\beta[dV(x)/dx]}{1-\kappa \beta V(x)}, \tag{10}$$

the stationary-state solution of F-P equation (8) can be written in the form of power-law distribution by

$$\rho_\kappa(x) = \frac{1}{Z_{\kappa-x}}[1-\kappa \beta V(x)]_+^{1/\kappa}, \tag{11}$$

where $[y]_+ = y$ for $y>0$, and zero otherwise; $Z_{\kappa-x}$ is a normalization constant that is determined by

$$Z_{\kappa-x} = \int dx [1-\kappa \beta V(x)]_+^{1/\kappa}. \tag{12}$$

As $\kappa \to 0$, all above recover to the forms with B-G distributions. Actually, it is clear that there is a set of stochastic dynamics in the Langevin equation with a whole set of $K(x)$ and $D(x)$ that give rise to the power-law distributions as the nonequilibrium stationary-state solutions of the corresponding F-P equation (8). As a specific example, if let $\rho_\nu(x)$ denote the stationary-state solution of F-P equation (8), and take $K(x) = -dV(x)/dx$ and $D(x) = D_x \nu [\rho_\nu(x)]^{\nu-1}$ with a parameter $\nu \neq 1$, we have

$$\frac{d}{dx}\left[\frac{dV(x)}{dx}\rho_\nu(x)\right] + D_x \frac{d^2}{dx^2}[\rho_\nu(x)]^\nu = 0. \tag{13}$$

If the above $K(x)$ and $D(x)$ are substituted into E.(10), and the parameter $\kappa$ is replaced by $(\nu-1)$, then the solution of the F-P equations (13) is directly written as

$$R_\nu(x) = \frac{1}{Z_{\nu-x}}[1-(\nu-1)\beta V(x)]_+^{1/(\nu-1)}, \tag{14}$$

where the normalization constant is gained by $Z_{v-x}^{v-1} = D_x v\beta$. As $v \to 1$, all above recover to the standard forms in conventional theory. It is novel to notice that this type of macroscopic state-dependent noise for $v \neq 1$ represents a kind of statistical feedback in the system away from equilibrium, and in this case an anomalous diffusion takes place, likely originating from the influence of environment on the microscopic dynamical processes.

The other type microscopic dynamics governed by the Langevin equation ably giving rise to macroscopic power-law distributions is for the momentum $p$, as a stochastic variable,

$$\frac{dp}{dt} = -\gamma\, p + \eta_p(t), \tag{15}$$

with

$$\langle \eta_p(t) \rangle = 0, \quad \langle \eta_p(t)\eta_p(t') \rangle = 2D(p)\delta(t-t'), \tag{16}$$

where $\gamma$ is a friction coefficient, and the noise $\eta_x(t)$ is Gaussian distribution, with zero mean and the delta-correlated. $D(p)$ can be a function of $p$ in general. Eq.(15) is a linear Langevin equation whose Gaussian stationary solution is shown to be changed by the multiplicative noise constituted due to the fluctuation in the friction coefficient $\gamma$, leading to the power-law form of Tsallis q-distribution [29]. But here we are to touch upon the other cases.

The F-P equation corresponding to the above Langevin equation (15) can be written for the noise-averaged probability distribution $\rho(p,t)$ as

$$\frac{\partial}{\partial t}\rho(p,t) = \frac{\partial}{\partial p}[\gamma\, p\rho(p,t)] + \frac{\partial}{\partial p}\left[D(p)\frac{\partial}{\partial p}\rho(p,t)\right]. \tag{17}$$

In the same way, similar to the condition (10), it is ready to find that if the following condition is satisfied for the function $D(p)$ and a given parameter $\kappa \neq 0$,

$$D(p) = \frac{m\gamma}{\beta}[1 - \kappa\, \beta p^2/2m], \tag{18}$$

the stationary-state solution of F-P equation (17) is found to be the following power-law form:

$$\rho_\kappa(p) = \frac{1}{Z_{\kappa-p}} \left[1 - \kappa\,\beta p^2/2m\right]_+^{1/\kappa}, \tag{19}$$

where the normalization constant $Z_{\kappa-p}$ reads

$$Z_{\kappa-p} = \int dp \left[1 - \kappa\,\beta p^2/2m\right]_+^{1/\kappa}. \tag{20}$$

There is also a novel example that if let $\rho_\nu(p)$ be the stationary-state solution of F-P equation (17), and take $D(p) = D_p \nu [\rho_\nu(p)]^{\nu-1}$ with a parameter $\nu \neq 1$, we get

$$\frac{d}{dp}[\gamma\, p\rho_\nu(p)] + D_p \frac{d^2}{dp^2}[\rho_\nu(p)]^\nu = 0. \tag{21}$$

After the above $D(p)$ is substituted into E.(18), and the parameter $\kappa$ is replaced by $(\nu-1)$, the solution of F-P equation (21) is written directly with the power-law form,

$$\rho_\nu(p) = \frac{1}{Z_{\nu-p}} \left[1 - (\nu-1)\beta p^2/2m\right]_+^{1/(\nu-1)}, \tag{22}$$

where the normalization constant is gained by $Z_{\nu-p}^{\nu-1} = D_p \nu \beta / m\gamma$.

There are a lot of power-law behaviors that have been observed and studied in various fields of science and technology, in which more complicated stochastic dynamical origins still need to be explored. In our situations where the equation of motion for the coordinate is modeled by the Langevin equation (4) and (5), and for the momentum by the Langevin equation (15) and (16), the complicated or anomalous diffusions in coordinate and momentum space, respectively, may play important roles to give rise to the power-law distributions. We mention that the F-P equations (13) and (21) have been the object of diverse recent and previous studies [30-33] and works elsewhere. If the parameter $\kappa$ in the expressions (11) and (19) is replaced by $(1-q)$, and also the parameter $\nu$ in the expressions (14) and (22) by $(2-q)$, it is picked up that these power-law distributions exactly become Tsallis $q$-distributions [23].

**III. THE TST RATES FOR THE SYSTEMS WITH POWER-LAW DISTRIBUTIONS**

We take the reaction processes (1) as a general example of the model to investigate the TST rate for power-law distributions. The rate equations, Eq.(2) and Eq.(3), for the reaction processes are taken into account. Let us follow the standard line [28] to generalize TST to the nonequilibrium system with the power-law stationary distributions. We consider a system with $n$ degrees of freedom $i = 1, 2, \ldots, n$, with coordinates $\{x_i\}$ and momenta $\{p_i\}$. The volume element of phase space is denoted by $d\Omega_n = dx_1 dx_2 \ldots dx_n dp_1 dp_2 \ldots dp_n$. An $n$-dimensional reaction system may be divided into one "reaction coordinate" and $n-1$ non-reactive coordinates. The reaction coordinate is denoted by $x$. Let us take a dividing surface to separate phase space into two regions $B$ and $A$, conventionally referred to as the reactant and product states. At the dividing surface, we take the reaction coordinate $x=0$, and $x>0$ if $x \in A$ and $x<0$ if $x \in B$. Thus the indicator function of $A$ can be expressed as the step function $\theta(x)$, defined by

$$\theta(x) = \begin{cases} 1, & \text{if } x > 0 \\ 0, & \text{otherwise} \end{cases} \tag{23}$$

Similarly, the indicator function of $B$ is $\theta(-x)$. If the phase space distribution function at time $t$ is $\rho(\{x_i\},\{p_i\},t)$, the instantaneous concentration of $A$, $N_A(t)$, can be defined as the ensemble average of $\theta(x)$, namely,

$$N_A(t) = \iiint_{\Omega_{n-1}} dx dp d\Omega_{n-1} \theta(x) \rho(\{x_i\},\{p_i\},t), \tag{24}$$

and $N_B(t)$ of $B$ as the ensemble average of $\theta(-x)$. And it is determined that

$$\begin{aligned} \frac{dN_A(t)}{dt} &= \iiint_{\Omega_{n-1}} dx dp d\Omega_{n-1} [\hat{L}\theta(x)] \rho(\{x_i\},\{p_i\},t) \\ &= \iiint_{\Omega_{n-1}} dx dp d\Omega_{n-1} \frac{p}{m} \delta(x) \rho(\{x_i\},\{p_i\},t), \end{aligned} \tag{25}$$

where $\hat{L}$ is the Liouville operator, and $\delta(x)$ is a delta function. Then, the instantaneous rate of $N_A(t)$ by transitions from $A$ to $B$ is expressed, by inserting the indicator function $\theta(-p)$ in Eq.(25), as

$$\left(\frac{dN_A(t)}{dt}\right)_{A \to B} = \iiint_{\Omega_{n-1}} dx dp d\Omega_{n-1} \frac{p}{m} \delta(x) \theta(-p) \rho(\{x_i\},\{p_i\},t). \tag{26}$$

Correspondingly, the rate from $B$ to $A$ is

$$\left(\frac{dN_A(t)}{dt}\right)_{B \to A} = \iiint_{\Omega_{n-1}} dx dp d\Omega_{n-1} \frac{p}{m} \delta(x) \theta(p) \rho(\{x_i\},\{p_i\},t). \tag{27}$$

It is frequently observed that the relaxation dynamics of nonequilibrium systems has not to be governed always by an exponential law, but in many situations it is by the non-exponential law or the power-law [4-21]. Thermal equilibrium is of eminent importance for the understanding of many processes in physics and chemistry, but in many other cases of complex systems far away from equilibrium, such as for example in living matter, fluxes of energy, matter and information prevent a system from approaching a thermal equilibrium state. Consequently, temperature is no longer uniform but may be with a distribution. Time and space dependent structure may then persist in the asymptotic long-time behavior of such a nonequilibrium system. Power-law distributions become a type of metastable states when the system reaches at an asymptotically stationary nonequilibrium. Due to the persistent exchanges of energy, matter and information between the system and its environment, as well as transformation among these physical quantities by chemical reactions taking place in the system, each particle is not free and it is always "feeling" the influences from the interactions with other particles in the system and the environment. Hence both entropy and energy in the processes are nonextensive or pseudoadditive. If the interactions might be modeled by a potential, e.g. $\varphi(x)$, the nonextensivity would be characterized in terms of a given parameter different from unity, e.g. $\nu$, referred to as measuring the degree of nonextensivity, by the relation [24-26]:

$$\nu - 1 \sim -\frac{dT}{dx} \bigg/ \frac{d\varphi(x)}{dx}, \tag{28}$$

where $T$ is temperature. It is clear that the parameter $\nu$ is different from unity if and only if the temperature gradient is not equal to zero; hence it represents a nonequilibrium stationary-state of the system away from thermal equilibrium.

Now we return to our discussions of the rate theory. For an autonomous nonequilibrium system, a stationary probability distribution will be approached asymptotically at long times. To investigate the transition rates between the regions $A$ and $B$ of the system under the condition away from thermal equilibrium, we assume that the system asymptotically reaches in the metastable states with stationary power-law distributions, and the power-law distributions can be expressed in the homologous forms to Eq.(14) and Eq.(22), which have been investigated in the generalized statistical mechanics [22, 23].

Let us consider a general many-body system with the Hamiltonian,

$$H = \sum_{i=1}^{n} \frac{p_i^2}{2m_i} + V(x_1, x_2, ..., x_n), \tag{29}$$

where the potential function is expressed as two-point interactions,

$$V(x_1, x_2, ..., x_n) = \sum_{i<j}^{n} u(|x_i - x_j|). \tag{30}$$

Then the full asymptotic stationary-state distribution of the nonequilibrium system is defined by the power-law $\nu$-distribution as

$$\rho_\nu(\{x_i\}, \{p_i\}) = \frac{1}{Z_\nu}\left[1 - (\nu-1)\beta H_\nu\right]_+^{1/(\nu-1)}, \tag{31}$$

where $\nu-1$ can be interchanged by $\kappa$ and $1-q$, thus is consistency with the formularies in previous section and some generalized statistical mechanics. The Hamiltonian quantity $H_\nu$ is $\nu$-pseudoadditive. The normalization constant in (31) is determined by

$$Z_\nu = \int_{\Omega_n} d\Omega_n \left[1 - (\nu-1)\beta H_\nu\right]_+^{1/(\nu-1)} = Z_\nu^A + Z_\nu^B, \tag{32}$$

with the contributions from the individual regions $A$ and $B$, respectively,

$$Z_\nu^A = \int_{\Omega_n} d\Omega_n \theta(x)\left[1 - (\nu-1)\beta H_\nu\right]_+^{1/(\nu-1)}, \tag{33}$$

and

$$Z_\nu^B = \int_{\Omega_n} d\Omega_n \theta(-x)\left[1 - (\nu-1)\beta H_\nu\right]_+^{1/(\nu-1)}. \tag{34}$$

In a living matter and a chemical reaction system far away from equilibrium, both entropy and energy are nonextensive or pseudoadditive due to the presence of persistent exchanges of energy, matter and information between the nonequilibrium system and its environments as well as by the interactions among particles inside the system. This kind of pseudoadditive characteristics has been studied in some generalized statistics made for power-law distributions, e.g. in nonextensive statistical mechanics [23, 34]. The relation between $H$ and $H_\nu$ [35, 36] is expressed as

$$1 - (\nu-1)\beta H_\nu = \prod_{i=1}^{n}\left[1-(\nu-1)\beta\frac{p_i^2}{2m_i}\right]_+ \cdot \prod_{i<j}^{n}[1-(\nu-1)\beta u(|x_i - x_j|)]_+ , \qquad (35)$$

so that the distribution function (31) can be written as

$$\rho_\nu(\{x_i\},\{p_i\}) = \frac{1}{Z_\nu}\prod_{i=1}^{n}\left[1-(\nu-1)\beta\frac{p_i^2}{2m}\right]_+^{1/(\nu-1)} \cdot \prod_{i<j}^{n}\left[1-(\nu-1)\beta u(|x_i - x_j|)\right]^{1/(\nu-1)} . \qquad (36)$$

The stationary-state concentration in region $A$ is $N_A = Z_\nu^A / Z_\nu$, and in region $B$, $N_B = Z_\nu^B / Z_\nu$. The local nonequilibrium distribution can be considered as the full stationary-state distribution weighted on either side by the actual amount of $A$ and $B$ that are present at time $t$. So the local nonequilibrium distribution in region $A$ is written as

$$\rho_\nu(\{x_i\},\{p_i\},t) = \frac{N_A(t)}{N_A}\rho_\nu(\{x_i\},\{p_i\}) . \qquad (37)$$

Accordingly, by substituting Eq.(37) into Eq.(26), we obtain

$$\left(\frac{dN_A(t)}{dt}\right)_{A\to B} = \iiint_{\Omega_{n-1}} dxdpd\Omega_{n-1}\frac{p}{m}\delta(x)\theta(-p)\rho_\nu(\{x_i\},\{p_i\})\frac{N_A(t)}{N_A}, \qquad (38)$$

which can be rewritten as the rate equation such as in Eq.(2), namely

$$\left(\frac{dN_A(t)}{dt}\right)_{A\to B} = -k_{AB}N_A(t) . \qquad (39)$$

Thereupon we find the reaction rate constant, given by

$$k_{AB} = -\frac{1}{N_A}\iiint_{\Omega_{n-1}} dxdpd\Omega_{n-1}\frac{p}{m}\delta(x)\theta(-p)\rho_\nu(\{x_i\},\{p_i\})$$

$$= \frac{1}{Z_\nu^A}\iiint_{\Omega_{n-1}} dxdpd\Omega_{n-1}\frac{p}{m}\delta(x)\theta(p)[1-(\nu-1)\beta H_\nu]_+^{1/(\nu-1)} . \qquad (40)$$

Or, equivalently, it can be written as

$$k_{AB} = \frac{\int_{\Omega_n} d\Omega_n \delta(x)\upsilon\theta(\upsilon)[1-(\nu-1)\beta H_\nu]_+^{1/(\nu-1)}}{\int_{\Omega_n} d\Omega_n \theta(x)[1-(\nu-1)\beta H_\nu]_+^{1/(\nu-1)}} , \qquad (41)$$

where $\upsilon$ is the velocity. In the same way, one has the rate equation from $B$ to $A$,

$$\left(\frac{dN_A(t)}{dt}\right)_{B\to A} = k_{BA} N_B(t), \tag{42}$$

and then we find the rate constant,

$$k_{BA} = \frac{\int_{\Omega_n} d\Omega_n \delta(x)\upsilon\theta(\upsilon)[1-(\nu-1)\beta H_\nu]_+^{1/(\nu-1)}}{\int_{\Omega_n} d\Omega_n \theta(-x)[1-(\nu-1)\beta H_\nu]_+^{1/(\nu-1)}}. \tag{43}$$

As expected, the TST rate constants for the thermal equilibrium assumption with B-G distribution are recovered in the limit $\nu \to 1$ in Eqs.(41) and (43). Thus we obtain a generalization of the TST rate constants to a nonequilibrium Hamiltonian system with the power-law distribution. In order to derive more specific expressions of the rate constants for the power-law distribution, we here present the following three cases for further discussions.

### A. The rate constants for a one-dimension Hamiltonian system

For one dimension system $n=1$, the Hamiltonian reads $H = p^2/2m + V(x)$, then from E.(35) and Eq.(36) one has the relation,

$$H_\nu = \frac{p^2}{2m} + V(x) + (\nu-1)\beta V(x)\frac{p^2}{2m}, \tag{44}$$

and the stationary-state distribution function,

$$\rho_\nu(x,p) = \frac{1}{Z_\nu}[1-(\nu-1)\beta p^2/2m]_+^{1/(\nu-1)}[1-(\nu-1)\beta V(x)]_+^{1/(\nu-1)}. \tag{45}$$

So the rate constant (41) becomes

$$k_{AB} = \frac{1}{Z_\nu^A} \iint dxdp\, \delta(x)\frac{p}{m}\theta(p)\rho_\nu(x,p,)$$

$$= \frac{1}{Z_{\nu-x}^A Z_{\nu-p}^A}[1-(\nu-1)\beta V(0)]_+^{1/(\nu-1)} \int_0^\infty dp\, \frac{p}{m}[1-(\nu-1)\beta p^2/2m]_+^{1/(\nu-1)}, \tag{46}$$

where the normalization constants have to be calculated in the two cases as follows, namely $0 < \nu < 1$ and $\nu > 1$, respectively. Particularly, for $\nu > 1$, the cutoff conditions must be taken into account.

For $0 < \nu < 1$, we have

$$Z_{\nu-x}^{A} = \int_{0}^{\infty} dx \left[1 - (\nu-1)\beta V(x)\right]_{+}^{1/(\nu-1)}, \tag{47}$$

and

$$Z_{\nu-p}^{A} = \int_{-\infty}^{\infty} dp \left[1 - (\nu-1)\beta p^2/2m\right]_{+}^{1/(\nu-1)}. \tag{48}$$

And for $\nu>1$, due to the cutoff condition $V(x) \leq 1/(\nu-1)\beta$, there is a value, $x=x_a$, that should be determined by $V(x_a) = max\ V(x) \leq 1/(\nu-1)\beta$, and also there is a maximum of $p$, $p_{max}$, that is $|p| \leq p_{max} = \sqrt{2m/\beta(\nu-1)}$, so that

$$Z_{\nu-x}^{A} = \int_{0}^{x_a} dx \left[1 - (\nu-1)\beta V(x)\right]_{+}^{1/(\nu-1)}, \tag{49}$$

and

$$Z_{\nu-p}^{A} = \int_{-p_{max}}^{p_{max}} dp \left[1 - (\nu-1)\beta p^2/2m\right]_{+}^{1/(\nu-1)}. \tag{50}$$

After completing the integrals of Eqs. (48) and (50), we obtain

$$Z_{\nu-p}^{A} = \sqrt{\frac{2m}{\beta}} \cdot \begin{cases} (\nu-1)^{-1/2} B\left(\frac{1}{2}, \frac{1}{\nu-1}+1\right), & \nu>1, \\ (1-\nu)^{-1/2} B\left(\frac{1}{2}, \frac{1}{\nu-1}-\frac{1}{2}\right), & 0<\nu<1, \end{cases} \tag{51}$$

where $B(a, b)$ is the beta function. The average velocity in Eq.(46) also needs to be calculated in the two cases, but the results in the both cases are the same: $<p/m>= 1/\beta\nu$. Substituting the above results into Eq.(46), we find the rate constants

$$k_{AB} = \frac{\sqrt{(1-\nu)/2m\beta}}{\nu B(\frac{1}{2}, \frac{1}{1-\nu}-\frac{1}{2})} \cdot \frac{\left[1-(\nu-1)\beta V(0)\right]_{+}^{1/(\nu-1)}}{\int_{0}^{\infty} dx \left[1-(\nu-1)\beta V(x)\right]_{+}^{1/(\nu-1)}}, \text{ for } 0<\nu<1, \tag{52}$$

and

$$k_{AB} = \frac{\sqrt{(\nu-1)/2m\beta}}{\nu B(\frac{1}{2}, \frac{1}{\nu-1}+1)} \cdot \frac{\left[1-(\nu-1)\beta V(0)\right]_{+}^{1/(\nu-1)}}{\int_{0}^{x_a} dx \left[1-(\nu-1)\beta V(x)\right]_{+}^{1/(\nu-1)}}, \text{ for } \nu>1. \tag{53}$$

It is clear that the TST rate expression with B-G distributions is recovered by the limit $\nu \to 1$ in above Eqs.(52) and (53), where the calculation in the limit $\nu \to 1$ for the beta function is performed by transforming it into gamma functions and then using the asymptotic formula for gamma functions [37]: $\lim_{|z|\to\infty} [z^{-\alpha}\Gamma(z+\alpha)/\Gamma(z)]=1$ (the calculations hereinafter for the beta and gamma functions are made in the same way). Thus, we receive a generalization of the TST rate constant to a one-dimension nonequilibrium system with the power-law distributions.

## B. The rate constants for an *n*-dimension Hamiltonian system

When we calculate the TST rate constants for an *n*-dimension system, we mention again that an *n*-dimensional reaction system has one "reaction coordinate" and $(n-1)$ non-reactive coordinates. Hereby, the numerator of Eq.(41) for the rate constant $k_{AB}$ is calculated by

$$\int_{\Omega_n} d\Omega_n \left\{ \delta(x) \upsilon \theta(\upsilon) \prod_{i=1}^{n} \left[1-(\nu-1)\beta p_i^2/2m_i\right]_+^{1/(\nu-1)} \cdot \prod_{i<j}^{n} \left[1-(\nu-1)\beta u(|x_i-x_j|)\right]_+^{1/(\nu-1)} \right\}$$

$$= \int_0^\infty dp \frac{p}{m} \left[1-(\nu-1)\beta p^2/2m\right]_+^{1/(\nu-1)} \cdot \prod_{i=2}^{n} \int_{-\infty}^{\infty} dp_i \left[1-(\nu-1)\beta \frac{p_i^2}{2m_i}\right]_+^{1/(\nu-1)}$$

$$\cdot \prod_{j=2}^{n} \left[1-(\nu-1)\beta u(|x_j|)\right]_+^{1/(\nu-1)} \cdot \prod_{\substack{i<j \\ i\neq 1}}^{n} \int_{-\infty}^{\infty} dx_i \left[1-(\nu-1)\beta u(|x_i-x_j|)\right]_+^{1/(\nu-1)}. \quad (54)$$

Correspondingly, the denominator of Eq.(41) for $k_{AB}$ is

$$\int_{\Omega_n} d\Omega_n \left\{ \theta(x) \prod_{i=1}^{n} \left[1-(\nu-1)\beta p_i^2/2m_i\right]_+^{1/(\nu-1)} \cdot \prod_{i<j}^{n} \left[1-(\nu-1)\beta u(|x_i-x_j|)\right]_+^{1/(\nu-1)} \right\}$$

$$= \int_{-\infty}^{\infty} dp \left[1-(\nu-1)\beta p^2/2m\right]_+^{1/(\nu-1)} \cdot \prod_{j=2}^{n} \int_0^{\infty} dx \left[1-(\nu-1)\beta u(|x-x_j|)\right]_+^{1/(\nu-1)}$$

$$\cdot \prod_{i=2}^{n} \int_{-\infty}^{\infty} dp_i \left[1-(\nu-1)\beta p_i^2/2m_i\right]_+^{1/(\nu-1)} \cdot \prod_{\substack{i<j \\ i\neq 1}}^{n} \int_{-\infty}^{\infty} dx_i \left[1-(\nu-1)\beta u(|x_i-x_j|)\right]_+^{1/(\nu-1)}. \quad (55)$$

So the rate constant $k_{AB}$, Eq.(41), reads

$$k_{AB} = \frac{\int_0^\infty dp(p/m)\left[1-(\nu-1)\beta p^2/2m\right]_+^{1/(\nu-1)}}{\int_{-\infty}^{\infty} dp \left[1-(\nu-1)\beta p^2/2m\right]_+^{1/(\nu-1)}} \cdot \frac{\prod_{j=2}^{n}\left[1-(\nu-1)\beta u(|x_j|)\right]_+^{1/(\nu-1)}}{\prod_{j=2}^{n} \int_0^\infty dx\left[1-(\nu-1)\beta u(|x-x_j|)\right]_+^{1/(\nu-1)}}. \quad (56)$$

The calculations of the prefactors in Eq.(56) also have to be performed in the two cases, namely $0 < \nu < 1$ and $\nu > 1$, respectively. And, for $\nu > 1$, the cutoff conditions should be taken into account. Finally, we derive the TST rate constants expressed with the power-law forms:

$$k_{AB} = \frac{\sqrt{(1-\nu)/2m\beta}}{\nu B(\frac{1}{2}, \frac{1}{1-\nu} - \frac{1}{2})} \cdot \frac{\prod_{j=2}^{n}\left[1-(\nu-1)\beta u(|x_j|)\right]_+^{1/(\nu-1)}}{\prod_{j=2}^{n} \int_0^\infty dx\left[1-(\nu-1)\beta u(|x-x_j|)\right]_+^{1/(\nu-1)}}, \text{ for } 0<\nu<1, \quad (57)$$

and

$$k_{AB} = \frac{\sqrt{(\nu-1)/2m\beta}}{\nu B(\frac{1}{2}, \frac{1}{\nu-1}+1)} \cdot \frac{\prod_{j=2}^{n}\left[1-(\nu-1)\beta u(|x_j|)\right]_+^{1/(\nu-1)}}{\prod_{j=2}^{n}\int_0^{x_a} dx\left[1-(\nu-1)\beta u(|x-x_j|)\right]_+^{1/(\nu-1)}}, \text{ for } \nu>1, \quad (58)$$

where due to the cutoff conditions, there is a value, $x=x_a$, that should be determined by $u(|x_a - x_j|) = \max u(|x - x_j|) \le 1/(\nu-1)\beta$, and at the same time $u(|x_j|)$ also should be cut by the constraint $u(|x_j|) \le 1/(\nu-1)\beta$. It is also clear that the TST rate constant for a $n$-dimension system with the B-G distribution is recovered by taking the limit $\nu \to 1$ in Eqs.(57) and (58). Thus, we obtain a generalization of the TST rate constants to the $n$-dimension nonequilibrium system with the power-law distributions.

### C. An generalized Arrhenius rate for power-law distributions

We take a typical application as an example to derive a generalized Arrhenius rate. In the chemical reaction system under consideration, the two regions $A$ and $B$ are associated with minima of the potential energy and are divided by a high barrier where the potential energy has a saddle point. Assume that in the neighborhood of the bottom of region $A$, located at $\{x_{ia}\}$, the potential energy is diagonal and expanded as a harmonic function,

$$V(\{x_i\}) = V_a + \frac{1}{2}m\omega_2(x-x_a)^2 + \frac{1}{2}\sum_{i=2}^{n} m_i \omega_i^2 (x_i - x_{ia})^2 + \cdots \quad (59)$$

where at the bottom, the potential has a minimum $V_a$. In the neighborhood of the saddle point, located at $\{x_i=0\}$, the potential energy is expanded as

$$V(\{x_i\}) = V_s - \frac{1}{2}m\omega^2 x^2 + \frac{1}{2}\sum_{i=2}^{n} m_i \omega_i^2 x_i^2 + \cdots \quad (60)$$

where the potential energy has a maximum $V_s$. Substitute Eq.(59) and Eq.(60) into the expression of $k_{AB}$, Eq.(41), then we find that the numerator of Eq.(41) is

$$\left[1-(\nu-1)\beta V_s\right]_+^{1/(\nu-1)} \cdot \int_{-\infty}^{\infty} dx \delta(x)\left[1-(\nu-1)\beta \tfrac{1}{2}m\omega^2 x^2\right]_+^{1/(\nu-1)}$$

$$\cdot \int_0^{\infty} dp \frac{p}{m}\left[1-(\nu-1)\beta p^2/2m\right]_+^{1/(\nu-1)} \cdot \prod_{i=2}^{n}\int_{-\infty}^{\infty} dp_i \left[1-(\nu-1)\beta \frac{p_i^2}{2m_i}\right]_+^{1/(\nu-1)}$$

$$\cdot \prod_{i=2}^{n} \int_{-\infty}^{\infty} dx_i \left[1 - (\nu - 1)\beta m_i \omega_i^2 x_i^2 / 2\right]_+^{1/(\nu-1)}, \tag{61}$$

and correspondingly, the denominator of Eq.(41) is

$$\left[1 - (\nu-1)\beta V_a\right]_+^{1/(\nu-1)} \cdot \int_0^\infty dx \left[1 - (\nu-1)\beta \tfrac{1}{2} m\omega^2 (x - x_a)^2\right]_+^{1/(\nu-1)}$$

$$\cdot \int_0^\infty dp \frac{p}{m} \left[1 - (\nu-1)\beta p^2 / 2m\right]_+^{1/(\nu-1)} \prod_{i=2}^{n} \int_{-\infty}^{\infty} dp_i \left[1 - (\nu-1)\beta \frac{p_i^2}{2m_i}\right]_+^{1/(\nu-1)}$$

$$\cdot \prod_{i=2}^{n} \int_{-\infty}^{\infty} dx_i \left[1 - (\nu-1)\beta \tfrac{1}{2} m_i \omega_i^2 (x_i - x_{ia})^2\right]_+^{1/(\nu-1)}. \tag{62}$$

So the generalized TST rate constant, Eq.(41), becomes

$$k_{AB} = \frac{\int_0^\infty dp(p/m)\left[1 - (\nu-1)\beta p^2/2m\right]_+^{1/(\nu-1)}}{\int_{-\infty}^\infty dp \left[1 - (\nu-1)\beta p^2/2m\right]_+^{1/(\nu-1)}}$$

$$\cdot \frac{\left[1 - (\nu-1)\beta V_s\right]_+^{1/(\nu-1)}}{\left[1 - (\nu-1)\beta V_a\right]_+^{1/(\nu-1)} \cdot \int_0^\infty dx\left[1 - (\nu-1)\beta \tfrac{1}{2} m\omega^2 (x - x_a)^2\right]_+^{1/(\nu-1)}}, \tag{63}$$

where the integral

$$\int_0^\infty dx \left[1 - (\nu-1)\beta \tfrac{1}{2} m\omega^2 (x - x_a)^2\right]_+^{1/(\nu-1)} = \sqrt{\frac{2}{\beta m\omega^2}} \int_{-x_a\sqrt{\beta m\omega^2/2}}^{\infty} dy \left[1 - (\nu-1)y^2\right]_+^{1/(\nu-1)}. \tag{64}$$

The probability distribution function in the integral must be zero at $y \to \pm\infty$. In the case of lower temperature, $\beta$ is sufficiently large so that the lower limit of the integral, $-x_a\sqrt{\beta m\omega^2/2}$, may be regarded as infinite in this integral. Consequently, it is found that the generalized TST rat constants are

$$k_{AB} = \frac{1-\nu}{\nu} \left[\frac{\Gamma\left(\tfrac{1}{1-\nu}\right)}{\Gamma\left(\tfrac{1}{1-\nu} - \tfrac{1}{2}\right)}\right]^2 \frac{\omega}{2\pi} \cdot \frac{\left[1 - (\nu-1)\beta V_s\right]_+^{1/(\nu-1)}}{\left[1 - (\nu-1)\beta V_a\right]_+^{1/(\nu-1)}}, \text{ for } 0<\nu<1, \tag{65}$$

and

$$k_{AB} = \frac{\nu-1}{\nu} \left[\frac{\Gamma\left(\tfrac{1}{\nu-1} + \tfrac{3}{2}\right)}{\Gamma\left(\tfrac{1}{\nu-1} + 1\right)}\right]^2 \frac{\omega}{2\pi} \cdot \frac{\left[1 - (\nu-1)\beta V_s\right]_+^{1/(\nu-1)}}{\left[1 - (\nu-1)\beta V_a\right]_+^{1/(\nu-1)}}, \text{ for } \nu>1. \tag{66}$$

As expected, when taking the limit $\nu \to 1$, they recover the TST rate in the familiar form with B-G exponential law,

$$k_{AB} = \frac{\omega}{2\pi} \cdot \exp[-\beta(V_s - V_a)], \tag{67}$$

which contains the familiar Arrhenius activation energy $(V_s - V_a)$ and the frequency factor $\omega/2\pi$. Thus we have received a generalized Arrhenius rate for a nonequilibrium system with the power-law distribution.

## IV. CONCLUSIONS

In conclusion, we investigate the transition state theory for a system away from thermal equilibrium when the system asymptotically reaches in a nonequilibrium stationary-state with the power-law distributions.

We firstly deal with the stochastic dynamics for the reaction coordinate and momentum, which is modeled by the Langevin equations and the corresponding Fokker-Planck equations, and then we study under what conditions the stochastic dynamics will give rise to the stationary power-law distributions. It is derived that if the condition Eq.(10) for the two functions $K(x)$ and $D(x)$ is satisfied for a given potential $V(x)$ and an any parameter $\kappa \neq 0$, the Langevin equation (4) for the coordinate $x(t)$ gives rise to the power-law distributions with the form such as Eq.(11). Eq.(4) is quite a general Langevin equation with arbitrary function $K(x)$, very similar to a strong friction or an overdamped motion of interacting particles in a potential $V(x)$ if $K(x)$ is associated with the potential by $K(x) = -dV(x)/dx$. As a specific case, it is shown that an anomalous diffusion, i.e. $D(x) = D_x \nu [\rho_\nu(x)]^{\nu-1}$ with an any given parameter $\nu \neq 1$, in the stochastic dynamics plays an important role to give rise to the stationary power-law distribution. In the homologous way, it is obtained that if the condition Eq.(18) for the function $D(p)$ is satisfied for an any given parameter $\kappa \neq 0$, the Langevin equation (15) for the momentum $p(t)$ gives rise to the power-law distributions with the form such as Eq.(19). It is also shown that as a specific case of the condition (18), an anomalous diffusion in the momentum space, i.e. $D(p) = D_p \nu [\rho_\nu(p)]^{\nu-1}$ with a given index $\nu \neq 1$, in the stochastic dynamics plays an important role to give rise to the stationary power-law distribution.

In the investigation of the generalized TST for a nonequilibrium system with the power-law distributions, we consider a general many-body Hamiltonian system. It is assumed that the system far away from equilibrium has not to relax to a thermal equilibrium state with the B-G distribution, but asymptotically approaches to a nonequilibrium stationary-state with the power-law distributions. Thus, following the standard line of TST we obtain a generalization of TST rates made suitable for a nonequilibrium system with the power-law distribution, given by Eq.(41) and Eq.(43) for positive and reverse reaction processes. Furthermore, we derive more specific expressions of the generalized TST rate constants for a one-dimension and an *n*-dimension cases for a nonequilibrium Hamiltonian system, given by Eqs.(52)-(53) and Eqs.(57)-(58), respectively. Finally, we consider the potential function as a harmonic approximation at the saddle point of the high barrier and the minimum of the potential bottom, respectively, thus we receive a generalized Arrhenius rate, given by Eqs.(65)-(66), for the nonequilibrium system with the power-law distribution.